# Implicit Lyapunov Control for the Quantum Liouville Equation

Shuang Cong [*,a], Fangfang Meng [a], Jianxiu Liu [a]

[*] Corresponding author. E-mail: scong@ustc.edu.cn
[a] Department of Automation, University of Science and Technology of China, Hefei, China, 230027

**Abstract**

A quantum system whose internal Hamiltonian is not strongly regular or/and control Hamiltonians are not full connected, are thought to be in the degenerate cases. In this paper, convergence problems of the multi-control Hamiltonians closed quantum systems in the degenerate cases are solved by introducing implicit function perturbations and choosing an implicit Lyapunov function based on the average value of an imaginary mechanical quantity. For the diagonal and non-diagonal target states, respectively, control laws are designed. The convergence of the control system is proved, and an explicit design principle of the imaginary mechanical quantity is proposed. By using the proposed method, the multi-control Hamiltonians closed quantum systems in the degenerate cases can converge from any initial state to an arbitrary target state unitarily equivalent to the initial state. Finally, numerical simulations are studied to verify the effectiveness of the proposed control method.

## 1. Introduction

In the last three decades, the quantum control theory developed rapidly. One of the main goals in quantum control theory is to develop a series of systematic methods for the control of quantum systems. There are many quantum control methods: quantum optimal control (Schmidt *et al*., 2011), adiabatic control (Boscain *et al*., 2012), quantum control method based on the Lyapunov stability theorem (Grigoriu, 2012; Wen & Cong, 2011), optimal Lyapunov-based quantum control (Hou, *et al*., 2012) and so on. By using the quantum Lyapunov control, control systems are at least stable. For this method, the convergence of control systems is a research focus, which is analyzed based on LaSalle's invariance principle (LaSalle & Lefschetz, 1961). According to LaSalle's invariance principle, as $t \to \infty$, any state trajectory will converge to the largest invariant set in the set $E$ in which the states satisfy that the first order derivative of the Lyapunov function equals zero. In fact, the set $E$ contains not only the target state but also other states which make the system may converge to other states. Thus, in order to enable the control system converge to the target state, the main idea is to add restrictions to make the set $E$ as small as possible. In recent years, research results on the convergence of the control system by using the Lyapunov control method are as follows:

I) Consider the Schrödinger equation $i\hbar|\dot{\psi}(t)\rangle = (H_0 + \sum_{k=1}^{r} H_k u_k(t))|\psi(t)\rangle$, where $|\psi(t)\rangle$ is the quantum state vector, $H_0$ is the internal Hamiltonian, $H_k, (k=1,\cdots,r)$ are control Hamiltonians, and $u_k(t), (k=1,\cdots,r)$ are control laws. For the target state $|\psi_f\rangle$ being an eigenstate, the conditions which make the control system converge to the target state are (Kuang & Cong, 2008):

i) The internal Hamiltonian is strongly regular, i.e., $\omega_{i'j'} \neq \omega_{lm}, (i',j') \neq (l,m), i',j',l,m \in \{1,2,\cdots,N\}$, where $\omega_{lm} = \lambda_l - \lambda_m$ represents the Bohr frequency (transition frequency) between the energy levels $\lambda_l$ and $\lambda_m$, $\lambda_l$ is the *l*-th eigenvalue of $H_0$ corresponding to the eigenstate $|\phi_l\rangle$.

ii) (a) For the Lyapunov function based on the state distance $V = 1/2\left(1 - |\langle\psi_f|\psi\rangle|^2\right)$ or state error $V = 1/2\langle\psi - \psi_f|\psi - \psi_f\rangle$, the condition is: all the eigenstates different from the target state are directly coupled to the target state, i.e., for $|\phi_i\rangle \neq |\psi_f\rangle$, there exists at least a $k$ such that $\langle\phi_i|H_k|\psi_f\rangle \neq 0$, or (b) For the Lyapunov function based on the average value of an imaginary mechanical quantity $V = \langle\psi|P|\psi\rangle$, where $P$ is an imaginary mechanical quantity, the condition is: any two eigenstates are coupled directly, i.e, for $|\phi_i\rangle \neq |\phi_j\rangle$, there exists at least a $k$ such that $\langle\phi_i|H_k|\phi_j\rangle \neq 0$.

II) Consider the quantum Liouville equation $i\hbar\dot{\rho}(t) = [H_0 + \sum_{k=1}^{r} H_k u_k(t), \rho(t)]$, where $\rho(t)$ is the density operator. For the target state being a diagonal matrix, the convergence conditions are (Wang & Schirmer, 2010; Kuang & Cong, Sep. 2010 & Feb. 2010): i) The internal Hamiltonian is strongly regular, and ii) The control Hamiltonians are full connected, i.e., $\forall j \neq l$, for $k = 1, \cdots, r$, there exists at least a $(H_k)_{jl} \neq 0$, where $(H_k)_{jl}$ is the (j,l)-th element of $H_k$.

In general, if the control systems satisfy the conditions mentioned above, they are called ideal systems and in the non-degenerate cases. However, many practical systems do not satisfy these conditions. They are so called in the degenerate cases. For the degenerate cases, people utilized a modified control design based on an "implicit" Lyapunov function so that the control system can converge to an arbitrary eigenstate from any pure state for the Schrödinger equation (Beauchard et al., 2007; Zhao et al., 2009 & 2012; Meng, Cong & Kuang, 2012). However, few works focus on the problem of convergence to a superposition state or a mixed state.

We propose a solution to this problem in this paper. The main contribution of this paper is to make the closed quantum system governed by the quantum Liouville equation in the degenerate cases can converge from an arbitrary initial state to an arbitrary target state unitarily equivalent to the initial state by using the implicit Lyapunov quantum control. In order to complete this control task, our method is to introduce implicit function perturbations into the control laws and choose an implicit Lyapunov function based on the average value of an imaginary mechanical quantity. Based on LaSalle's invariance principle, analyze the convergence of the control system and seek the convergence conditions. Then analyze how to make these convergence conditions be satisfied, and propose the explicit design principle of the imaginary mechanical quantity.

The remainder of this paper is arranged as follows: in Section 2, the model of the control system and control objective are described. In Section 3, the Lyapunov function is choosed, and the control laws are designed. In Section 4, the convergence of the control system is analyzed and proved. How to make the convergence conditions of the control system be satisfied is analyzed. And the explicit design principle of the imaginary mechanical quantity is proposed and proved. Then a design method for the non-diagonal target state cases is proposed. In Section 5, numerical simulations on a 3-level system are done to verify the effectiveness of the proposed method. Some concluding remarks are drawn in Section 6.

## 2. Description of Problem

Consider the $N$-level closed quantum control system governed by the following Quantum Liouville Equation:

$$i\dot{\rho}(t) = [H_0 + \sum_{k=1}^{r} H_k u_k(t), \rho(t)], \rho_0 = \rho(0), (\hbar = 1) \quad (1)$$

where $\rho(t)$ is the density operator, $H_0$ is the internal Hamiltonian, $H_k, (k = 1, \cdots, r)$ are control Hamiltonians, and $u_k(t), (k = 1, \cdots, r)$ are control laws.

In order to solve the convergence problem of the control system in the degenerate cases, we introduce perturbations $\gamma_k(t)$ into the control laws. Thus Eq. (1) becomes

$$i\dot{\rho}(t) = [H_0 + \sum_{k=1}^{r} H_k(\gamma_k(t) + v_k(t)), \rho(t)], \rho_0 = \rho(0) \quad (2)$$

where $\gamma_k(t)$ are the control laws to solve the convergence problem for the degenerate cases, and $\gamma_k(t) + v_k(t) = u_k(t), (k = 1, \cdots, r)$ are the total control laws.

The control objective is to design control laws $u_k(t) = \gamma_k(t) + v_k(t)$ so as to make the control system depicted by (2) can completely transfer from any initial state $\rho_0$ to an arbitrary desired target state $\rho_f = U^H \rho_0 U, U \in SU(N)$.

## 3. Design of control laws

In this paper, the Lyapunov quantum control method based on the average value of an imaginary mechanical quantity is used. The so-called imaginary mechanical quantity means that it is a linear Hermitian operator to be designed, and maybe not a physical observable. In order to solve the convergence problem for the degenerate cases, perturbations $\gamma_k(t), (k = 1, \cdots, r)$ are introduced into the control laws, and the specific Lyapunov function is selected as:

$$V(\rho) = tr(P_{\gamma_1, \cdots, \gamma_r} \rho) \quad (3)$$

where $P_{\gamma_1, \cdots, \gamma_r} = f(\gamma_1(t), \cdots, \gamma_r(t))$ is a functional of $\gamma_k(t)$ and is positive definite.

After introducing perturbations $\gamma_k(t)$, $H_0 + \sum_{k=1}^{r} H_k \gamma_k(t)$ can be regarded as the new internal Hamiltonian of the control system. In order to facilitate understanding the basic idea of this method, we describe the system in the eigenbasis of $(H_0 + \sum_{k=1}^{r} H_k \gamma_k)$. Assume the eigenvalues and eigenstates of $(H_0 + \sum_{k=1}^{r} H_k \gamma_k)$ are $\lambda_{n,\gamma_1,\cdots,\gamma_r}$ and $|\phi_{n,\gamma_1,\cdots,\gamma_r}\rangle$, $1 \leq n \leq N$, respectively. Set $U_1 = (|\phi_{1,\gamma_1,\cdots,\gamma_r}\rangle, \cdots, |\phi_{N,\gamma_1,\cdots,\gamma_r}\rangle)$, then the control system in the eigenbasis of $(H_0 + \sum_{k=1}^{r} H_k \gamma_k)$ is

$$i\dot{\hat{\rho}}(t) = [(\hat{H}_0 + \sum_{k=1}^{r}\hat{H}_k\gamma_k(t)) + \sum_{k=1}^{r}\hat{H}_k v_k(t), \hat{\rho}(t)] \qquad (4)$$

where $\hat{\rho} = U_1^H \rho U_1, \hat{H}_0 = U_1^H H_0 U_1, \hat{H}_k = U_1^H H_k U_1$.

Accordingly, the target state $\rho_f$ will become $\hat{\rho}_f = U_1^H \rho_f U_1$ which is also a functional of $\gamma_k(t)$.

The design idea of $\gamma_k(t)$ is as follows: 1) The perturbations are designed to satisfy the strongly regular and full connected conditions so that the control system can converge toward $\hat{\rho}_f$ by designing appropriate control laws; 2) $\gamma_k(t), (k=1,\cdots,r)$ need converge to zero, and their convergent speed must be slower than that of the control sytem to $\hat{\rho}_f$ to make $\gamma_k(t)$ take effect; 3) $\gamma_k(\rho_f)=0$ must hold to make the control system be asymptotically stable at the target state $\rho_f$.

For the non-degenerate cases, Kuang and Cong proposed the restriction $V(\rho_f) < V(\rho_0) < V(\rho_{other})$ to make the system converge to the target state $\rho_f$ from the initial state $\rho_0$, where $\rho_{other}$ represents any other state in the set $E = \{\rho | \dot{V}(\rho)=0\}$ except the target state (Kuang & Cong, Sep. 2010 & Feb. 2010). Without this restriction, the system state trajectory is possible to evolve to $\rho_{other}$, at which all the control laws vanish because of $\dot{V}(\rho)=0 \Leftrightarrow u_k(t)=0, (k=1,\cdots,r)$. Then the system will keep staying in $\rho_{other}$. Namely, the control laws fail to manipulate the control system converge to the target state. In order to make the state trajectory not stay in $\rho_{other}$ until it reaches the target state for the degenerate cases, we can design that for $k=1,\cdots,r$, all the perturbations $\gamma_k(t) = 0$ holds only in the target state, i.e., 1) $\gamma_k(\rho_f)=0, (k=1,\cdots,r)$, and 2) for $\rho \neq \rho_f$, there exists at least one $k$ such that $\gamma_k(\rho) \neq 0$. Thus the restriction only need to be $V(\rho_f) < V(\rho_{other})$ for the degenerate cases.

According to the analysis mentioned above, let us design $\gamma_k(t), (k=1,\cdots,r)$.

In the Lyapunov control, since evolution of the system's state relies on the decrease of the Lyapunov function, we design $\gamma_k(t)$ be a monotonically increasing functional of $V(t)$:

$$\gamma_k(\rho) = C_k \cdot \theta_k(V(\rho) - V(\rho_f)) \qquad (5)$$

where $C_k \geq 0$, and for $k=1,\cdots,r$, there exists at least a $C_k > 0$. And $\theta_k(\cdot)$ satisfies $\theta_k(0)=0$, $\theta_k(s) > 0$, $\theta_k'(s) > 0$ for every $s > 0$. The existence of $\gamma_k(t)$ can be depicted by Lemma 1.

**Lemma 1:** If $C_k = 0$, $\gamma_k(\rho) = 0$. Else if $C_k > 0$, $\theta_k \in C^\infty(R^+;[0,\gamma_k^*]), k=1,\cdots,r$ ($\gamma_k^*$ is a positive constant) satisfy $\theta_k(0) = 0$, $\theta_k(s) > 0$, $\theta_k'(s) > 0$ for every $s > 0$, and $|\theta_k'| < 1/(2C^*C_k)$, $C^* = 1+C$, $C = \max\{\|\partial P_{\gamma_1,\cdots,\gamma_r}/\partial \gamma_k\|_{m_1}, (k=1,\cdots,r)\}$, then for every $\rho$, there is a unique $\gamma_k \in C^\infty (\gamma_k \in [0,\gamma_k^*])$ satisfying $\gamma_k(\rho) = C_k \cdot \theta_k(tr(P_{\gamma_1,\cdots,\gamma_r}\rho) - tr(P_{\gamma_1,\cdots,\gamma_r}\rho_f)), (k=1,\cdots,r)$.

**Proof:** Assume $P_{\gamma_1,\cdots,\gamma_r}$ are analytic functions of the perturbations $\gamma_k(\rho) \in [0,\gamma_k^*], (k=1,\cdots,r)$. $\partial P_{\gamma_1,\cdots,\gamma_r}/\partial\gamma_k, (k=1,\cdots,r)$ are bounded on $[0,\gamma_k^*]$, thus $C < \infty$. By (3) and (5), the derivative of $\theta_k$ on $\gamma_k$ is

$$\partial \theta_k / \partial \gamma_k = \theta_k' tr(\partial P_{\gamma_1,\cdots,\gamma_r}/\partial \gamma_k (\rho - \rho_f)) \qquad (6)$$

Define

$$F_k(\gamma_1,\cdots,\gamma_r,\rho) = \gamma_k - C_k \cdot \theta_k(tr(P_{\gamma_1,\cdots,\gamma_r}\rho) - tr(P_{\gamma_1,\cdots,\gamma_r}\rho_f)) \qquad (7)$$

where $F_k(\gamma_1,\cdots,\gamma_r,\rho), (k=1,\cdots,r)$ are regular. For a fixed $\rho$,

$$F_k(\gamma_1(\rho),\cdots,\gamma_r(\rho),\rho) = 0, (k=1,\cdots,r) \qquad (8)$$

holds. By (6), one can obtain

$$\partial F_k / \partial \gamma_k = 1 - C_k \theta_k' tr(\partial P_{\gamma_1,\cdots,\gamma_r}/\partial \gamma_k \cdot (\rho - \rho_f)) \qquad (9)$$

Some deductions show that

$$|tr(\partial P_{\gamma_1,\cdots,\gamma_r}/\partial \gamma_k \cdot (\rho - \rho_f))| \leq 2\|\partial P_{\gamma_1,\cdots,\gamma_r}/\partial \gamma_k\|_{m_1} \qquad (10)$$

According to the given condition, one can have

$$|C_k \theta_k' tr(\partial P_{\gamma_1,\cdots,\gamma_r}/\partial \gamma_k \cdot (\rho - \rho_f))| < 1 \qquad (11)$$

Then

$$\partial F_k(\gamma_1(\rho),\cdots,\gamma_r(\rho),\rho)/\partial \gamma_k \neq 0 \qquad (12)$$

holds. Thus according to the implicit Theorem (Krantz & Parks, 2002), Lemma 1 is proved. □

Then let us design $v_k(t)$. The basic idea is to design the control laws $v_k(t)$ such that the time derivative of the selected Lyapunov function $\dot{V}(t) \leq 0$ holds. For the sake of simplicity, set $\gamma_k(t)=0$ for some $k$, and other $\gamma_k(t)$ are equal, denoted by $\gamma(t)$, i.e., set

$$\gamma_k(t)=\gamma(t), C_k=1, k=k_1,\cdots,k_m;$$
$$C_k=0, k\neq k_1,\cdots,k_m (1\leq k_1,\cdots,k_m\leq r) \quad (13)$$

Correspondingly, (3) can be rewritten as $V(\rho)=tr(P_\gamma\rho)$, where $P_\gamma$ is a functional of $\gamma(t)$. By (2), we can obtain the time derivative of the selected Lyapunov function as follows:

$$\dot{V}=-itr([P_\gamma, H_0+\sum_{n=k_1}^{k_m}H_n\gamma(t)]\rho)-$$
$$i\sum_{k=1}^{r}v_k(t)tr([P_\gamma, H_k]\rho)+\dot{\gamma}tr((\partial P_\gamma/\partial\gamma)\rho) \quad (14)$$

The sign of the first term in the right-hand side of (14) is difficult to determine. By setting $[P_\gamma, H_0+\sum_{n=k_1}^{k_m}H_n\gamma(t)]=0$, this term is eliminated. Then (14) becomes

$$\dot{V}=-i\sum_{k=1}^{r}v_k(t)tr([P_\gamma, H_k]\rho)+\dot{\gamma}tr((\partial P_\gamma/\partial\gamma)\rho) \quad (15)$$

Equation (15) contains the time derivative of the implicit function perturbation $\dot{\gamma}(t)$ which needs to be eliminated. By (5) and (13), we can obtain the time derivative of the implicit function perturbation $\dot{\gamma}(t)$ as:

$$\dot{\gamma}(t)=(i\theta'\sum_{k=1}^{r}v_k tr([P_\gamma, H_k]\rho))\Big/(\theta'tr((\partial P_\gamma/\partial\gamma)(\rho-\rho_f))\text{-}1) \quad (16)$$

Substituting (16) into (15), one has

$$\dot{V}=-\frac{1+\theta'tr((\partial P_\gamma/\partial\gamma)\rho_f)}{1-\theta'tr((\partial P_\gamma/\partial\gamma)(\rho-\rho_f))}\sum_{k=1}^{r}itr([P_\gamma, H_k]\rho)v_k(t) \quad (17)$$

According to (11) and (13), one can obtain $|\theta'tr((\partial P_\gamma/\partial\gamma)\rho_f)|<1/2$, then $(1+\theta'tr((\partial P_\gamma/\partial\gamma)\rho_f))\big/(1\text{-}\theta'tr((\partial P_\gamma/\partial\gamma)(\rho-\rho_f)))>0$ holds. In order to ensure $\dot{V}(t)\leq 0$, $v_k(t), (k=1,\cdots,r)$ are designed as:

$$v_k(t)=K_k f_k\big(itr([P_\gamma, H_k]\rho)\big), (k=1,\cdots,r) \quad (18)$$

where $K_k$ is a constant and $K_k>0$, and $y_k=f_k(x_k), (k=1,2,\cdots,r)$ are monotonic functions through the coordinate origin and in the first quadrant and the third quadrant of the plane $x_k-y_k$.

## 4. Convergence analysis

In this section, the convergence is analyzed based on LaSalle's invariance principle (LaSalle & Lefschetz, 1961). According to LaSalle's invariance principle, if the Lyapunov function $V(t)$ satisfies $V(t)>0$, $\dot{V}(t)\leq 0$, as $t\to\infty$, any trajectory will converge to the largest invariant set in $E=\{\rho\,|\,\dot{V}(\rho)=0\}$. After the above analysis, the convergence of the control system can be depicted by Theorem 1.

**Theorem 1**: Consider the control system depicted by (2) with control laws $\gamma_k(t)$ defined by Lemma 1, Eq. (5) and (13), and $v_k(t)$ defined by (18), if the control system satisfies: i) $\omega_{l,m,\gamma_1,\cdots,\gamma_r}\neq\omega_{i,j,\gamma_1,\cdots,\gamma_r}, (l,m)\neq(i,j), i,j,l,m\in\{1,2,\cdots,N\}$, $\omega_{l,m,\gamma_1,\cdots,\gamma_r}=\lambda_{l,\gamma_1,\cdots,\gamma_r}-\lambda_{m,\gamma_1,\cdots,\gamma_r}$, where $\lambda_{l,\gamma_1,\cdots,\gamma_r}$ is the $l$-th eigenvalue of $(H_0+\sum_{k=1}^{r}H_k\gamma_k)$ corresponding to the eigenvector $|\phi_{l,\gamma_1,\cdots,\gamma_r}\rangle$; ii) $\forall j\neq l$, for $k=1,\cdots,r$, there exists at least a $(\hat{H}_k)_{jl}\neq 0$, where $(\hat{H}_k)_{jl}$ is the $(j,l)$-th element of $\hat{H}_k=U_1^H H_k U_1$ with $U_1=\big(|\phi_{1,\gamma_1,\cdots,\gamma_r}\rangle,\cdots,|\phi_{N,\gamma_1,\cdots,\gamma_r}\rangle\big)$; iii) $[P_\gamma, H_0+\sum_{n=k_1}^{k_m}H_n\gamma(t)]=0, 1\leq k_1,\cdots,k_m\leq r$; iv) For any $l\neq j, (1\leq l,j\leq N)$, $(P_\gamma)_{ll}\neq(P_\gamma)_{jj}$ holds, where $(P_\gamma)_{ll}$ is the $(l,l)$-th element of $P_\gamma$, then the control system will converge toward the largest invariant set in $E=\{\rho\,|\,\rho_{lj}=0; 1\leq l,j\leq N\}$, where $\rho_{lj}$ is the $(l,j)$-th element of $\rho$.

**Proof**:

Without loss of generality, assume that for $t\geq t_0, (t_0\in R)$, $\dot{V}=0$ is satisfied. By (17) and (18), one can obtain

$$\dot{V}=0 \Leftrightarrow tr([P_\gamma, H_k]\rho)=0 \Leftrightarrow v_k(t)=0 \quad (19)$$

As $\dot{V}=0$, $\gamma$ are constants, denoted by $\bar{\gamma}$. By property of the trace, (19) can be written as

$$\dot{V}=0 \Leftrightarrow tr([\hat{P}_{\bar{\gamma}}, \hat{H}_k]\hat{\rho})=0 \Leftrightarrow v_k(t)=0 \quad (20)$$

where $\hat{P}_{\bar{\gamma}}=U_1^H P_{\bar{\gamma}} U_1$. Set $\hat{\rho}_{t_0}=\hat{\rho}(t_0)$, the solution of Eq. (4) with $\gamma_k(t)$ defined by Eq. (5), Eq. (13), $\gamma=\bar{\gamma}$ and $v_k(t)=0$ is

$$\hat{\rho}(t)=e^{-i(\hat{H}_0+\sum_{n=k_1}^{k_m}\hat{H}_n\bar{\gamma})(t-t_0)}\hat{\rho}_{t_0}e^{i(\hat{H}_0+\sum_{n=k_1}^{k_m}\hat{H}_n\bar{\gamma})(t-t_0)} \quad (21)$$

Thus $tr([\hat{P}_{\bar{\gamma}}, \hat{H}_k]\hat{\rho}) = 0$ can be written as

$$tr(e^{-i(\hat{H}_0 + \sum_{n=k_1}^{k_m} \hat{H}_n\bar{\gamma})(t-t_0)} \hat{\rho}_{t_0} e^{i(\hat{H}_0 + \sum_{n=k_1}^{k_m} \hat{H}_n\bar{\gamma})(t-t_0)} [\hat{P}_{\bar{\gamma}}, \hat{H}_k]) = 0 \quad (22)$$

By condition iii), $\hat{P}_{\bar{\gamma}}$ is a diagonal matrix. Therefore

$$\hat{P}_{\bar{\gamma}} = e^{-i(\hat{H}_0 + \sum_{n=k_1}^{k_m} \hat{H}_n\bar{\gamma})(t-t_0)} P_{\bar{\gamma}} e^{i(\hat{H}_0 + \sum_{n=k_1}^{k_m} \hat{H}_n\bar{\gamma})(t-t_0)}$$ holds. Substituting it into (22), one can obtain

$$tr(e^{i(\hat{H}_0 + \sum_{n=k_1}^{k_m} \hat{H}_n\bar{\gamma})(t-t_0)} \hat{H}_k e^{-i(\hat{H}_0 + \sum_{n=k_1}^{k_m} \hat{H}_n\bar{\gamma})(t-t_0)} [\hat{\rho}_{t_0}, \hat{P}_{\bar{\gamma}}]) = 0 \quad (23)$$

By $e^A B e^{-A} = \sum_{n=0}^{\infty} (1/n!)[A^{(n)}, B]$, one gets

$$\sum_{n=0}^{\infty} (1/n!)(i^n (t-t_0)^n) tr([(\hat{H}_0 + \sum_{n=k_1}^{k_m} \hat{H}_n\bar{\gamma})^{(n)}, \hat{H}_k][\hat{\rho}_{t_0}, \hat{P}_{\bar{\gamma}}]) = 0 \quad (24)$$

where

$$[(\hat{H}_0 + \sum_{n=k_1}^{k_m} \hat{H}_n\bar{\gamma})^{(n)}, \hat{H}_k] = \underbrace{[(\hat{H}_0 + \sum_{n=k_1}^{k_m} \hat{H}_n\bar{\gamma}), [(\hat{H}_0 + \sum_{n=k_1}^{k_m} \hat{H}_n\bar{\gamma}), \cdots, \hat{H}_k]]}_{n \; times}$$

Then gives

$$\sum_{j,l=1}^{N} \omega_{j,l,\bar{\gamma}}^n (\hat{H}_k)_{jl} ((\hat{P}_{\bar{\gamma}})_{ll} - (\hat{P}_{\bar{\gamma}})_{jj})(\hat{\rho}_{t_0})_{lj} = 0, (k=1,\cdots,r) \quad (25)$$

where $(\hat{P}_{\bar{\gamma}})_{ll}$ is the (l,l)-th element of $\hat{P}_{\bar{\gamma}}$. Set

$$\xi_k = \begin{bmatrix} (\hat{H}_k)_{12}((\hat{P}_{\bar{\gamma}})_{22} - (\hat{P}_{\bar{\gamma}})_{11})(\hat{\rho}_{t_0})_{21} \\ \vdots \\ (\hat{H}_k)_{(N-1)N}((\hat{P}_{\bar{\gamma}})_{NN} - (\hat{P}_{\bar{\gamma}})_{(N-1)(N-1)})(\hat{\rho}_{t_0})_{N(N-1)} \end{bmatrix}$$

(26a)

$$\Lambda = diag(\omega_{1,2,\bar{\gamma}}, \cdots, \omega_{N-1,N,\bar{\gamma}}) \quad (26b)$$

$$M = \begin{bmatrix} 1 & 1 & \cdots & 1 \\ \omega_{1,2,\bar{\gamma}}^2 & \omega_{1,3,\bar{\gamma}}^2 & \cdots & \omega_{N,N-1,\bar{\gamma}}^2 \\ \vdots & \vdots & \vdots & \vdots \\ \omega_{1,2,\bar{\gamma}}^{N(N-1)-2} & \omega_{1,3,\bar{\gamma}}^{N(N-1)-2} & \cdots & \omega_{N,N-1,\bar{\gamma}}^{N(N-1)-2} \end{bmatrix} \quad (26c)$$

For $n=0,2,4,\cdots$, (25) reads $M\Im(\xi_k) = 0$. For $n=1,3,5,\cdots$, (25) reads $M\Lambda\Re(\xi_k) = 0$. By condition i), $M$ and $\Lambda$ are nonsingular real matrices. One can obtain $\xi_k = 0$, i.e.,

$$(\hat{H}_k)_{jl}((\hat{P}_{\bar{\gamma}})_{ll} - (\hat{P}_{\bar{\gamma}})_{jj})(\hat{\rho}_{t_0})_{lj} = 0, (k=1,\cdots,r) \quad (27)$$

By condition ii), one can have

$$((\hat{P}_{\bar{\gamma}})_{ll} - (\hat{P}_{\bar{\gamma}})_{jj})(\hat{\rho}_{t_0})_{lj} = 0, (k=1,\cdots,r) \quad (28)$$

That is $[\hat{P}_{\bar{\gamma}}, \hat{\rho}_{t_0}] = 0, (k=1,\cdots,r)$. Denote $\rho_{t_0} = \rho(t_0)$, then $[P_{\bar{\gamma}}, \rho_{t_0}] = 0, (k=1,\cdots,r)$, i.e.,

$$((P_{\bar{\gamma}})_{ll} - (P_{\bar{\gamma}})_{jj})(\rho_{t_0})_{lj} = 0, (k=1,\cdots,r) \quad (29)$$

holds. By condition iv), one can get

$$(\rho_{t_0})_{lj} = 0 \quad (30)$$

Then Theorem 1 is proved based on the LaSalle's invariance principle. □

From Theorem 1, one can see that if conditions i)-iv) hold, the control system will converge to a diagonal matrix. Since the evolution of $\rho(t)$ is unitary in closed quantum systems, $\rho(t)$ for $t \geq 0$ are isospectral. Denote the eigenvalues of the initial state $\rho_0$ by $\lambda_{01}, \lambda_{02}, \cdots, \lambda_{0N}$, then the diagonal elements of the states in $E$ are various permutations of $\lambda_{01}, \lambda_{02}, \cdots, \lambda_{0N}$. Therefore $E$ has at most $N!$ elements.

In the evolution process of the system's states, as $\dot{V}(t) \leq 0$, the Lyapunov function is decreasing. In order to make the system converge to the target state, we need to design $P_{\gamma}$ to make $V(\rho_f)$ be smallest, i.e.,

$$V(\rho_f) < V(\rho_{other}) \quad (31)$$

where $\rho_{other}$ represents any other state in the set $E$ except the target state. With the condition (31), the state trajectory will not stay in $\rho_{other}$ until the target state $\rho_f$ is reached because there exists at least one non-zero control law in any $\rho_{other}$ to make the state trajectory evolve. Therefore for the target state being a diagonal matrix, if the control system satisfies the conditions in Theorem 1 and (31), the state trajectory can converge to the target state from an arbitrary initial state unitarily equivalent to the target state.

Next we'll analyze how to make these conditions be satisfied in detail. Conditions i) and ii) in Theorem 1 are associated with $H_0$, $H_k$, $(k=1,\cdots,r)$ and $\gamma_k(t)$. By appropriately designing $\gamma_k(t)$, these two conditions can be satisfied in most cases. Condition iii) means that $P_{\gamma}$ and $H_0 + \sum_{n=k_1}^{k_m} H_n \gamma(t)$ have the same eigenstates, denoted by $|\phi_{j,\gamma}\rangle, (j=1,\cdots,N)$. Design the eigenvalues of $P_{\gamma}$ be constant, denoted by $P_1, P_2, \cdots, P_N$, then $P_{\gamma}$ can be written as

$$P_\gamma = \sum_{j=1}^{N} P_j |\phi_{j,\gamma}\rangle\langle\phi_{j,\gamma}| \tag{32}$$

Generally, $H_0$ is a diagonal matrix whose eigenvectors are $[1,0,0,\cdots,0]^T, \cdots [0,0,0,\cdots,1]^T$. As perturbations $\gamma(t)$ are very small, if we design $P_l \neq P_j (\forall l \neq j; 1 \leq l, j \leq N)$ and make $P_1, P_2, \cdots, P_N$ not be close to each other, in general, condition iv) can be satisfied. Then let us analyze how to make (31) hold. The result is as follows:

**Theorem 2:** For the diagonal target state $\rho_f$, if $(\rho_f)_{ii} < (\rho_f)_{jj}, 1 \leq i,j \leq N$, design $(P_\gamma)_{ii} > (P_\gamma)_{jj}$; if $(\rho_f)_{ii} = (\rho_f)_{jj}, 1 \leq i,j \leq N$, design $(P_\gamma)_{ii} \neq (P_\gamma)_{jj}$; else if $(\rho_f)_{ii} > (\rho_f)_{jj}, 1 \leq i,j \leq N$, design $(P_\gamma)_{ii} < (P_\gamma)_{jj}$, then $V(\rho_f) < V(\rho_{other})$ holds.

**Proof:**
At first, some useful propositions are proposed as follows.

**Proposition 1:** If the diagonal elements of the diagonal target state $\{(\rho_f)_{11}, (\rho_f)_{22}, \cdots, (\rho_f)_{NN}\}$ arranged in a decreasing order, design $\{(P_\gamma)_{11}, (P_\gamma)_{22}, \cdots, (P_\gamma)_{NN}\}$ arranged in an increasing order, then $V(\rho_f) < V(\rho_{other})$ holds.

**Proof:** Denote any other state contained in the set $E$ as:
$$\rho_{other} = diag((\rho_f)_{11(\tau)}, (\rho_f)_{22(\tau)}, \cdots, (\rho_f)_{NN(\tau)}) \tag{33}$$

where $\{11(\tau), 22(\tau), \cdots, NN(\tau)\}$ is a permutation of $\{11, 22, \cdots, NN\}$.

The Lyapunov function $V(\rho) = tr(P_\gamma \rho)$ can be written as
$$V(\rho) = tr(P_\gamma \rho) = \sum_{j=1}^{N} (P_\gamma)_{jj} \rho_{jj} \tag{34}$$

Assume $(\rho_f)_{11} > (\rho_f)_{22} > \cdots > (\rho_f)_{NN} \geq 0$, and $0 < (P_\gamma)_{11} < (P_\gamma)_{22} < \cdots < (P_\gamma)_{NN}$.

For $N=2$,
$$\begin{aligned} V(\rho_f)_2 &- V(\rho_{other})_2 \\ &= ((P_\gamma)_{11} - (P_\gamma)_{22})((\rho_f)_{11} - (\rho_f)_{22}) < 0 \end{aligned} \tag{35}$$

where the subscript "2" in $V(\rho_f)_2$ and $V(\rho_{other})_2$ means $N=2$. Proposition 1 is true.

For $N=3$, one can also prove Proposition 1 is true.
Assume that Proposition 1 is true for $N-1$. Then

$$\begin{aligned} V(\rho_f)_{N-1} - V(\rho_{other})_{N-1} &= \sum_{j=1}^{N-1} (P_\gamma)_{jj} ((\rho_f)_{jj} - (\rho_f)_{jj(\tau)}) \\ &= \sum_{j=1}^{N-1} ((P_\gamma)_{jj(\tau)} - (P_\gamma)_{jj})(\rho_f)_{jj(\tau)} < 0 \end{aligned} \tag{36}$$

For $N$,
$$\begin{aligned} V(\rho_f)_N - V(\rho_{other})_N &= \sum_{j=1}^{N-1} ((P_\gamma)_{jj(\tau)} - (P_\gamma)_{jj})(\rho_f)_{jj(\tau)} \\ &+ ((P_\gamma)_{NN(\tau)} - (P_\gamma)_{NN})(\rho_f)_{NN(\tau)} \end{aligned} \tag{37}$$

By (36) and $0 < (P_\gamma)_{11} < (P_\gamma)_{22} < \cdots < (P_\gamma)_{NN}$, one can get
$$V(\rho_f)_N - V(\rho_{other})_N < 0 \tag{38}$$

Proposition 1 is proved. □

**Proposition 2:** If the diagonal elements of the diagonal target state $\{(\rho_f)_{11}, (\rho_f)_{22}, \cdots, (\rho_f)_{NN}\}$ arranged in a non-decreasing order with $(\rho_f)_{k_{11}k_{11}} = \cdots = (\rho_f)_{k_{1l_1}k_{1l_1}} < (\rho_f)_{k_{21}k_{21}} = \cdots = (\rho_f)_{k_{2l_2}k_{2l_2}} < \cdots < (\rho_f)_{k_{q1}k_{q1}} = \cdots = (\rho_f)_{k_{ql_q}k_{ql_q}}, 1 \leq k_{ij} \leq N, k_{11} = 1, k_{ql_q} = N$, design $\{(P_\gamma)_{11}, (P_\gamma)_{22}, \cdots, (P_\gamma)_{NN}\}$ as follows: $(P_\gamma)_{k_{11}k_{11}}, \cdots, (P_\gamma)_{k_{1l_1}k_{1l_1}} > \cdots > (P_\gamma)_{k_{q1}k_{q1}}, \cdots, (P_\gamma)_{k_{ql_q}k_{ql_q}} > 0$. Then, $V(\rho_f) < V(\rho_{other})$ holds.

**Proof:** Obviously, Proposition 2 is true for $N=2, 3$.
Assume that for $N-1$, Proposition 2 is true. Then (36) holds.

For $N$, if $(\rho_f)_{(N-1)(N-1)} < (\rho_f)_{NN}$, design $(P_\gamma)_{k_{11}k_{11}}, \cdots, (P_\gamma)_{k_{1l_1}k_{1l_1}} > \cdots > (P_\gamma)_{NN}$, then (38) holds.

If $(\rho_f)_{k_{q1}k_{q1}} = (\rho_f)_{k_{q2}k_{q2}} = \cdots = (\rho_f)_{NN}$, then $NN(\tau) \neq k_{q1}k_{q1} \neq \cdots \neq k_{q(l_q-1)}k_{q(l_q-1)}$ in (37). Design $(P_\gamma)_{k_{11}k_{11}}, \cdots, (P_\gamma)_{k_{1l_1}k_{1l_1}} > \cdots > (P_\gamma)_{k_{q1}k_{q1}}, \cdots, (P_\gamma)_{k_{ql_q}k_{ql_q}}$, then (38) holds. Proposition 2 is proved. □

Obviously, according to Proposition 1, Proposition 2 and (34), we can obtain Theorem 2. □

Although $(P_\gamma)_{ij}$ is not easy to construct due to its dependence on $\gamma_k(t)$, it can be designed in principle as follows: $P_\gamma$ is defined by (32), and if $(\rho_f)_{ii} < (\rho_f)_{jj}, 1 \leq i,j \leq N$, then $P_i > P_j$; if $(\rho_f)_{ii} = (\rho_f)_{jj}, 1 \leq i,j \leq N$, then $P_i \neq P_j$; else if $(\rho_f)_{ii} > (\rho_f)_{jj}, 1 \leq i, j \leq N$, then $P_i < P_j$, and design $P_1, P_2, \cdots, P_N$ not be close to each other. Then regulate parameters in accordance with the result of the experiment.

From the convergence analysis mentioned above, one can see that the proposed method can only ensure that the control system converge to the diagonal target state. When the target state is a superposition state or non-diagonal mixed state, further research needs to be done. Our design idea is to utilize a unitary transformation on the coordinate axes to transform the non-diagonal target state in the original coordinate system to a diagonal matrix in the new coordinate system. Correspondingly, the dynamical equation (2), the initial state and other physical quantities will also be transformed. Then design the control laws based on the methods proposed in Sections 3.

Assume that $\rho_f$ is transformed to a diagonal matrix $\tilde{\rho}_f$ by a unitary transformation $\tilde{\rho}_f = U_2^H \rho_f U_2$. Substituting $\rho = U_2 \tilde{\rho} U_2^H$ into the dynamical equation (2), gives

$$i\dot{\tilde{\rho}}(t) = [\tilde{H}_0 + \sum_{k=1}^{r} \tilde{H}_k(\gamma_k(t) + v_k(t)), \tilde{\rho}(t)], \tilde{\rho}_0 = \tilde{\rho}(0) \quad (39)$$

where $\tilde{H}_0 = U_2^H H_0 U_2, \tilde{H}_k = U_2^H H_k U_2$.

In this case, any state trajectory governed by (2) converging to the non-diagonal target state $\rho_f$ is equivalent to any state trajectory governed by (39) converging to the diagonal matrix $\tilde{\rho}_f$. According to the design of the control laws in Section 3, the selected Lyapunov function is $V(\tilde{\rho}) = tr(P_\gamma \tilde{\rho})$, $\gamma_k(\tilde{\rho}), k = 1, \cdots, r$ satisfy (13) and $\gamma(t) = \theta(tr(P_\gamma \tilde{\rho}) - tr(P_\gamma \tilde{\rho}_f))$, and $v_k(t) = K_k f_k(itr([P_\gamma, \tilde{H}_k]\tilde{\rho}))$, one can prove that the convergence (Theorem 1) and the design principle of the imaginary mechanical quantity (Theorem 2) also hold with every parameter and physical quantity changing accordingly.

## 5. Numerical simulations

In order to verify the effectiveness of the proposed method, a three-level system is considered in this section. And numerical simulations are done for a non-diagonal target state case.

Consider a three-level system with $H_0$ and $H_1$ as:

$$H_0 = \begin{bmatrix} 0.3 & 0 & 0 \\ 0 & 0.6 & 0 \\ 0 & 0 & 0.9 \end{bmatrix}, H_1 = \begin{bmatrix} 0 & 1 & 1 \\ 1 & 0 & 0 \\ 1 & 0 & 0 \end{bmatrix} \quad (40)$$

According to $H_0$ and $H_1$, the system is in the degenerate case.

Choose a superposition state $\rho_0 = \begin{bmatrix} 0.1667 & 0.2357 & 0.2887 \\ 0.2357 & 0.3333 & 0.4083 \\ 0.2887 & 0.4083 & 0.5000 \end{bmatrix}$ as the initial state, and another superposition state $\rho_f = \begin{bmatrix} 0.3333 & 0.4714 & 0 \\ 0.4714 & 0.6667 & 0 \\ 0 & 0 & 0 \end{bmatrix}$ as the target state.

According to the basic design idea introduced in Section 4, by unitary transformation $\tilde{\rho} = U_2^H \rho U_2$, $U_2 = \begin{bmatrix} -0.8165 & 0 & 0.57735 \\ 0.57735 & 0 & 0.8165 \\ 0 & 1 & 0 \end{bmatrix}$, $\rho_f$ is transformed to a diagonal matrix $\tilde{\rho}_f = diag\{0, 0, 1\}$. Accordingly, the control system in the new coordinate system is (39) with $r = 1$. According to the proposed design ideas, the control law is $u_1(t) = \gamma_1(t) + v_1(t)$. $\gamma_1(t)$ is designed as $\gamma_1(\rho) = M_1 \cdot (tr(P_{\gamma_1} \tilde{\rho}) - tr(P_{\gamma_1} \tilde{\rho}_f))$ and $v_1(t)$ is designed as: $v_1(t) = K_1(itr([P_{\gamma_1}, \tilde{H}_1]\tilde{\rho}))$, where $M_1$ is the proportional coefficient of $\gamma_1$ and $M_1 > 0$, and $P_{\gamma 1} = \sum_{j=1}^{3} P_j |\tilde{\phi}_{j,\gamma_1}\rangle$ in which $|\tilde{\phi}_{j,\gamma_1}\rangle$ is the $j$-th eigenvector of $\tilde{H}_0 + \tilde{H}_1 \gamma_1(t)$.

In the numerical simulations, according to the specific design principle of the imaginary mechanical quantity proposed in Section 4, design $P_1, P_2, P_3$ such that $P_1 \neq P_2 > P_3$ and not be close to each other. Denote the terminal time of numerical simulations as $t_f$, the principle of regulating $P_1, P_2, P_3$ is as follows: in general, if $(\tilde{\rho}(t_f))_{ii} < (\tilde{\rho}_f(t_f))_{ii}$, decrease $P_i$; else if $(\tilde{\rho}(t_f))_{ii} > (\tilde{\rho}_f(t_f))_{ii}$, increase $P_i$, where $(\tilde{\rho}(t_f))_{ii}$ and $(\tilde{\rho}_f(t_f))_{ii}$ are the $(i,i)$-th elements of $\tilde{\rho}(t_f)$ and $\tilde{\rho}_f(t_f)$, respectively. Sometimes this method dose not work because $P_i$ not only influence $\tilde{\rho}_{ii}$ ($(i,i)$-th element of $\tilde{\rho}$), but also influence other elements of $\tilde{\rho}$. The principle of regulating $K_1$ is: the larger $K_1$ is, the faster the system converges to the target state. But after $K_1$ exceeding a certain value, the transition probability will decrease, the control effect will deteriorate, and the system even may oscillate. The principle of regulating $M_1$ is: regulate $M_1$ be as small as possible on the premise that the perturbations take effect to make the control system exhibits a behavior of trending to the target state. After tuning the control parameters repeatedly and carefully, the control parameters are selected as: $M_1 = 0.1, K_1 = 0.25, P_1 = 1.5$, $P_2 = 2.1$, and $P_3 = 0.01$.

In the simulations, the time step size is set as 0.01 a.u., and the control duration is 30 a.u.. The results of numerical simulations are shown in Fig.1 and Fig.2. Fig.1 displays the evolution curves of $\rho_{11}$, $\rho_{12}$, $\rho_{22}$ and $\rho_{33}$. Fig.2 shows the designed control laws $v_1(t)$, $\gamma_1(t)$ and $u_1(t)$.

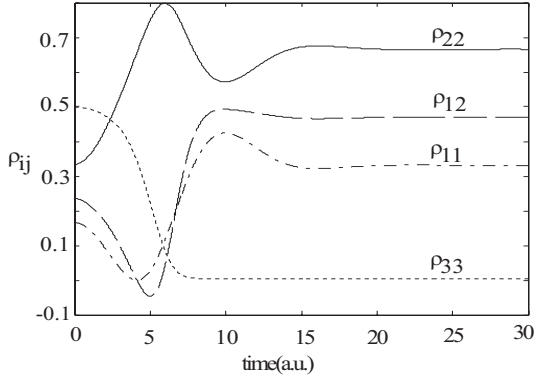

Fig. 1. The elements of density matrix $\rho$ of the control system

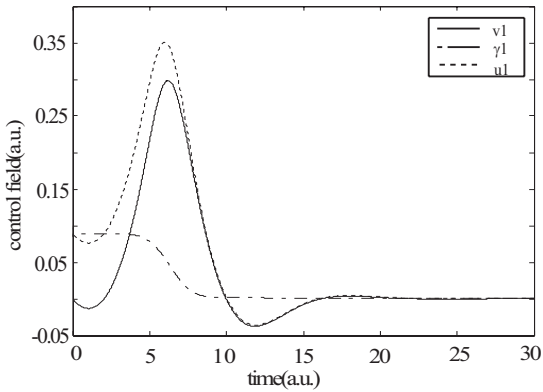

Fig. 2. Control fields of the control system

According to numerical results, the state at 30 a.u. is $\rho_{11} = 0.33069$, $\rho_{12} = 0.46921$, $\rho_{22} = 0.66576$ and $\rho_{33} = 0.0035519$, and the transition probability reaches $99.65\%$. So the proposed control method is effective.

## 6. Conclusion

In this paper, we have investigated the convergence for the closed quantum systems governed by the quantum Liouville equation. For the so-called degenerate cases where the internal Hamiltonian is not strongly regular or/and the control Hamiltonians are not full connected, by introducing implicit function perturbations and choosing an implicit Lapunov function based on the average vale of an imaginary mechanical quantity, an implicit Lyapunov control method has been proposed to complete the state transfer task from an arbitrary initial state to an arbitrary target state unitarily equivalent to the initial state. According to the LaSalle invariance principle, convergence of the control system has been analyzed and proved. The conditions for convergence have been anyalyzed, for which a specific design principle of the imaginary mechanical quantity also has been proposed. At last, the numerical results have demonstrated that the proposed control method in our work is correct and effective.

**Acknowledgements**

This work was supported partly by the National Key Basic Research Program under Grant No. 2011CBA00200, and the National Science Foundation of China under Grant No. 61074050.